\documentclass[lettersize,journal]{IEEEtran}
\usepackage{amsmath,amsfonts}
\usepackage{algorithm}
\usepackage{algpseudocode}
\usepackage{array}
\usepackage[caption=false,font=normalsize,labelfont=sf,textfont=sf]{subfig}
\usepackage{textcomp}
\usepackage{stfloats}
\usepackage{url}
\usepackage{verbatim}
\usepackage{graphicx}
\usepackage{cite}
\usepackage[subtle]{savetrees}
\usepackage[dvipsnames,tabl,xcdraw]{xcolor}
\usepackage{graphicx}

\begin{document}

\title{Assessment of the Center of Inertia and Regional Inertia with Load Contribution via a Fully Data-Driven Method}

\author{Lucas Lugnani,~\IEEEmembership{Student Member,~IEEE,} Mario R. A. Paternina,~\IEEEmembership{Member,~IEEE} Daniel Dotta,~\IEEEmembership{Member,~IEEE}
}



\maketitle

\begin{abstract}
This paper proposes a new comprehensive and fully data-driven methodology to estimate the center of inertia (COI) and the regional inertia, considering the displacement of the COI due to disturbances and load inertial contributions. The strategy uses the typicality-based data analysis (TDA) technique to detect the right pilot-bus that represents the COI. In the TDA, a compound of correlation and cosine similarities is implemented to approximate the actual distribution of the data and find the point (bus) closest to the mean which is elected as the pilot-bus. Then, the frequency response at the pilot-bus and the active power deviations are embedded into an autoregressive moving average exogenous input (ARMAX)-based approach to determine the regional inertia represented by an equivalent machine, whose inertia constant corresponds to the inertial contribution in the Region. 
The methodology is tested using the IEEE 68-bus benchmark test system and an adapted version with aggregated dynamical loads, corroborating the method effectiveness.
\end{abstract}


\section{Introduction}
Power systems are transitioning to renewable sustainable sources, and the role of inertia is becoming more critical to system frequency stability~\cite{milano2018foundations}. Wind and solar energy account for a continuously higher share of the generation fleet of power systems and are explored using converter connected generators (CCG). Thanks to the DC-link, CCGs are isolated from electromechanical phenomena in the system, like loss of generation and load, preventing their response capabilities which must be determined via control algorithms, unlike natural acting inertia from synchronous generators. Therefore, such phenomena are becoming more severe as the penetration of CCGs in power systems increases~\cite{miller12,arani14}. To appropriately deal with these kind of events in modern power systems, operators must have accurate estimates of the inertia for every Region to counteract their actions and ensure the system frequency stability. In that sense, the installation of wide-area monitoring systems (WAMS) in power systems in the last decades is beneficial, as it provides a great amount of data that enriches the situational awareness and decision-making of operators. From the literature, recent investigations have supported the inertia assessment using WAMS measurements.


\subsection{Literature Review}

Several works in the literature deal with the estimation of inertia of synchronous generators at their point of interconnection using synchrophasors~\cite{Wall14,gorbunov2019estimationGen,lugnani2020armax}. However, the assessment of the total system inertia or regional inertia imposes additional challenges. For instance, in~\cite{Wall14} an inertia estimation is conducted using WAMS without considering the COI displacement.
Other approaches address the COI's estimation~\cite{cepeda2014_realCOI1,milano2017rotor,zhao2018robust,you2020calculate,azizi2020local,gorbunov2022COIambient}, without discussing the regional inertia estimation. Likewise, some works tackle the regional inertia estimation to some extent classifying them by their distinct characteristics into the following categories: (\textit{i}) disturbance methods~\cite{Ashton15,wilson2019measuring,Zografos17,alshahrestani2018wams,schiffer2019online,phurailatpam2019measurement,yang2020data,makolo2021online,kerdphol2021determining}; (\textit{ii}) probing signal methods~\cite{zhang2017,tamrakar2020inertia}; and (\textit{iii}) ambient-signal methods~\cite{Wilson18,cai2019inertia,yang2020ambient,Wang2022InertiaAmbientOscillation,cui2020ambient,allella2020line,zeng2020onlineCOIambientAccess,baruzzi2021analysis}. 
As we are interested in regional inertia estimation, which entails larger systems, the occurrence of severe events is fairly greater than for a single machine, for instance, providing a reasonable amount of opportunities for estimation. It is also important to point out that severe disturbance methods provide reference values of estimation for the development of other two types of method, that is, probing and ambient signal methods. Hence, the choice of a disturbance method is advocated and we focus our investigation on disturbance methods reported in the literature.

In~\cite{Ashton15}, the authors rely on extensive WAMS measurements and event detection and selection using detrended fluctuation analysis (DFA) to monitor clusters of generators in the GB system. Frequency signals stemming from PMUs are filtered using a low-pass filter, and the power deviation of generators is estimated. Then, the ratio between a known power deviation and its estimate multiplied by the total inertia of synchronous generators produces the estimate of total system inertia, considering load contribution. The authors in~\cite{wilson2019measuring} perform a report on the effective inertia of the GB and Icelandic system, taking into account load contribution and using the swing equation. 
In~\cite{Zografos17}, the total inertia estimation is computed using WAMS measurements, particle swarm optimization (PSO), 
and load contributions that are conceived into an optimal formulation, where loads are modeled as voltage-dependent 
The method is validated at the Nordic57 test system. In~\cite{alshahrestani2018wams}, authors use frequency and active power measurements along with the knowledge of generators inertia to fit the frequency response of a disturbance using polynomial techniques. This method is tested in the IEEE 68-bus test system. In~\cite{schiffer2019online}, the assessment of the equivalent inertia is done using a first-order nonlinear aggregated power system model in combination with the recently proposed dynamic regressor and mixing (DREM). Where the equivalent machine inertia is estimated approximating the active power deviation through the power deviation caused by primary frequency control and the COI frequency response by a simple average of all generators frequency responses. In~\cite{phurailatpam2019measurement}, 
a polynomial fitting is performed over the swing equation and frequency measurements, 
demonstrating robustness to topology and location of disturbance. 
The method proposed in~\cite{yang2020data} uses dynamic mode decomposition (DMD) to extract the eigenvalues and eigenvectors, from which inertia is derived. This method does not require the COI 
knowledge, since it is based on inter-area electromechanical oscillations. However, the accuracy may be influenced by topology changes that shift the modes' frequencies. In~\cite{makolo2021online}, the equivalent inertia of a power system is estimated using the recursive least-squares (RLS) by fitting  oscillation data, where an initial model is estimated with a non-recursive system identification method. 
The authors in~\cite{kerdphol2021determining} estimate the inertia of areas from the 60Hz Japan system using frequency and rate of change of frequency measurements applying a frequency spectrum and performing mode shape analysis. Here, 
they find that PMUs further away from the COI of the system provide imprecise estimations due to the effect of inter-area oscillation.

It is evident in the literature that regional inertia estimation methods do not consider the proper estimation of the COI frequency~\cite{kerdphol2021determining,guo2022onlineHCOI} and the load contribution to the effective inertia, making simple assumptions. 
The assumption that loads contribute to the inertia response is well established in~\cite{ash15}, but overall ignored until recent years with the high penetration of CCG in modern systems. As these generators have small inertia constants and are isolated from the system by the converter interface, but are displacing synchronous generation, it becomes useful for system operators to acknowledge and estimate the load inertial contribution as an important resource for frequency stability. 

\subsection{Contribution}

This investigation proposes a disturbance based and data-driven method for the detection of regional COI pilot-bus, using a compound of the \textit{cosine} and \textit{correlation} distance metrics of frequency and active power signals at buses. This compound distance is processed by the TDA's features \cite{chow13Coherency} to approximate the probability distribution of the regional inertial responses and find the highest \textit{typicality} value, corresponding to the mean of the distribution, that is, the COI. 

Our proposal also estimates the effective regional inertia through a swing equation equivalent machine representation for each Region, where the Region tie-lines active power is used as input, whereas the pilot-bus frequency response as output, using ARMAX model identification. The estimations of the regional equivalent machine inertia are validated comparing with the actual inertia of the IEEE 68-bus simulated power system, in two scenarios: with or without load contributions. Where aggregated induction motors are added to the load buses of the system model. 

Thus, the primary contributions of this research are enclosed in the following: \textit{i)} a fully data-driven detection of the COI pilot-bus is achieved using only disturbance synchrophasor measurements; \textit{ii)} the sensibility to the load inertia impact in the detection of the COI pilot-bus is investigated; \textit{iii)} a fully data-driven regional equivalent inertia estimation $H_{est}$ is conducted using disturbance measurements and a variable order ARMAX-based identification model.

The remainder of the paper describes the representation of the COI per Region through an estimated pilot-bus and the effects of the load in the inertial response in Section~\ref{sec:proposal}. Section~\ref{sec:method} discloses the methodology for inertial estimation using pilot-bus frequency and tie-line active power signals and ARMAX-based identification. Section~\ref{sec:results} presents the validation of the methodology using the IEEE 68-bus NETS/NYPS test system and its modified version considering dynamical load representation. Finally, Section~\ref{sec:conclusion} summarizes the presented contributions regarding our proposed methodology and points out the future works related to the inertial response estimation in power systems.

\section{Fundamentals: Pilot-bus Representation and Inertial Estimation considering Load Contribution}\label{sec:proposal}

\subsection{Pilot-bus detection using TDA}\label{ssec:FundPilotBus}

The assumption that the Region is coherent, it is important for regional inertia estimation because the frequency response of buses within a single Region will present the same trend, that is, it will be unimodal. In this sense, note that \eqref{eq:fcoi} is a weighted average, that is:

\begin{equation}
    f_{coi}(t) = \frac{\sum_{n=1}^{N_g}f_{n}(t)H_{n} + \sum_{m=1}^{N_m}f_{m}(t)H_{m}}{\sum_{n=1}^{N_g}H_{n} + \sum_{m=1}^{N_m}H_{m}} = \frac{\sum_{i=1}^{N}x_{i}w_{i}}{\sum_{i=1}^{N}w_{i}}
    \label{eq:weightedfcoi}
\end{equation}

\noindent where $N_g$ is the number of generators plus synchronous condensers in the Region, $N_{m}$ is the number of load buses with motors connected to them. $f_{n}(t)$ and $H_{n}$ are respectively the frequency and inertia of each generator and synchronous condenser, $f_{m}(t)$ is the frequency of the transmission load bus to which a considerable amount of motors, i.e. a industrial district, is connected via a distribution system, and $H_{m}$ is the equivalent inertia of the motors connected at that load bus. On the right-hand side, $x_{i}$ denotes frequency measurements of every generator or motor bus, meanwhile $w_{i}$ symbolizes its respective inertia. Since $f_{coi}(t)$ is essentially virtual, it may not necessarily correspond to the frequency of any particular bus of the Region. Note also that, for non-generator and non-motor load buses its frequency ($f_{k}(t)$) is a function $g$ of the Region inertias and the admittance matrix ($Y_{R}$) of the Region~\cite{milano2016frequency}:


\begin{equation}
    f_{k}(t) \sim g(H_{[n,m]},Y_{R})
    \label{eq:freqfHY}
\end{equation}

Since $f_{coi}(t)$ is a virtual quantity, it may be arbitrarily close to any frequency in the Region. For example, a generator with inertia orders of magnitude that are higher than any others in the Region, or a bus which corresponds to the center of a symmetrical Region:

\begin{equation}
    f_{coi}(t) \approx f_{i}(t), \forall i \in N=\{n,m,i\}
    \label{eq:fcoiapprox}
\end{equation}

\noindent where $N$ is the set of all buses in the Region. We can represent the distance of the frequency response of each bus to the virtual $f_{coi}$ by a probability density function (pdf) of any type (e.g. Gaussian, colored, Weibull, etc). However, we do not have information regarding the type of distribution, nor precise knowledge of the weights (inertias) to calculate the mean of the distribution, but only synchrophasor measurements. Using TDA \cite{chow13Coherency} we are able to approximate the pdf of the Region, where we find the bus whose frequency $f_{i}(t)$ is closest to the COI frequency $f_{coi}(t)$, the mean of the Region's pdf~\cite{angelov2019empirical}.

To find the bus closest to the COI, we assume that the inertial response of each bus $f(t)$ is the first 2 seconds ($t_f$) of after disturbance ($t_0$), where no speed governor has had time to act. Additionally, $f(t)$ can be represented by the Euclidean norm $\beta$ of its frequency deviation ($\Delta f(t)$) with respect to the nominal frequency ($f_{0}$). To determine the closest bus to the mean ($f_{coi}(t)$), we calculate the norm between the frequency deviations of every pair of buses $k$ and $j$ $\beta(k,j)$ to indicate the closeness to the mean of the pdf. Then, the inertial response for bus $k$ with respect to bus $j$ can be represented by:

\begin{equation}
    \beta(k,j)=\sqrt{\sum_{t=t_{0}}^{t_{f}}[\Delta f_{k}(t) - \Delta f_{j}(t)]^2}
    \label{eq:EuclideanNormf}
\end{equation}

To find the pilot-bus that embodies the inertial response of the Region, 
the electrical power response deviation of every bus $\pi(k,j)$ with respect to every other bus $j$ is expressed in \eqref{eq:EuclideanNormP}. It is important to consider the relative norm between buses $k$ and $j$ for electrical power as a weighting factor, since generator and motor buses will present higher power deviations than other buses (transmission buses). This is due to their inertial content, which may deviate the mean of the distribution. For transmission buses, total power equals zero, so we convention that the power injected by the bus at the grid is considered, as in the generators case. Hence, for any bus $k$ with respect to any other bus $j$, the electrical power norm $\pi(k,j)$ will be:

\begin{equation}
    \pi(k,j)=\sqrt{\sum_{t=t_{0}}^{t_{f}}[\Delta Pe_{k}(t) - \Delta Pe_{j}(t)]^2}
    \label{eq:EuclideanNormP}
\end{equation}

\noindent where the power deviation $\Delta Pe_{k}(t)$ is calculated with respect to the value of the electrical power at $t=t_0$.

For every bus $k$ in the Region $R$ with $N$ buses, the inertial response of $k$ can be represented by a vector $2 \times N$, forming a point $\alpha(k)$ in the distribution space. For the TDA application, a distance metric in this space must be defined for the computation of the properties that will produce the approximation of the distribution's pdf. The choice of the metric for TDA must consider the physical aspects such as space and phenomena in question. Also, the metric may be compounded to consider relevant aspects that may be secluded to one given metric. For such purpose, a compound distance metric which takes into account the angle delay between buses (cosine metric - $d_{cos}$) and the linear coefficient of their distribution (correlation metric - $d_{corr}$) is used. Hence, the compound distance metric $\delta$ is given by:

\begin{equation}
    \delta(k,j) = d_{cos} + d_{corr} = \frac{\alpha(k)\cdot \alpha(j)}{\alpha(k)\times \alpha(j)} + \frac{cov(\alpha(k),\alpha(j))}{\sigma_{\alpha(k)}\sigma_{\alpha(j)}}
    \label{eq:TDAdist}
\end{equation}

\noindent where $cov(\alpha(k),\alpha(j))$ stands for the covariance between points $\alpha(k)$ and $\alpha(j)$, and $\sigma_{\alpha(k)}$ represents the standard deviation of $\alpha(k)$ and likewise for $\alpha(j)$.

Now, we can define the TDA properties as proposed in~\cite{Lugnani2021TDA}, where the aim is to produce \textit{typicality} values $\tau_{N}(\alpha(k))$, for each point $k$ indicating the distribution of the inertial response, according to algorithm~\ref{alg:TDA}.
\begin{algorithm}[!t]
	\caption{TDA implementation for COI pilot-bus detection.}
	\begin{algorithmic}[1]
		\State {\bf Input:} Let $\mathbf{\alpha}_{k}$, $k = 1,\dots N$ (data points) vector of scalar Euclidean norms between active power responses $\mathbf{\pi(k,j)}$ and frequencies responses $\mathbf{\beta(k,j)}$, with $N$ being the number of PMUs in Region $R$. 
		\State {\bf Output:} Pilot-bus that represents the COI of the Region $R$ with maximal typicality $\mathbf{\tau}_{N}^{*}$.
		\State {\bf Initialization:} $\mathbf{t}_{0}$, $\mathbf{t}_{f}$, set of correlation metrics $\delta_{k,j}$
		    \For{k=1,k++}
		        \For{j=1,j++}
		            \State $\delta_{k,j} \gets \frac{\alpha(k).\alpha(j)}{\alpha(k)\times \alpha(j)} + \frac{cov(\alpha_{k},\alpha_{j})}{\sigma_{k}\sigma_{j}} \hspace{0.5cm} \alpha_{k},\alpha_{j} \in {\alpha}_{N}$
	            \EndFor
		    \EndFor
	    \State {\bf TDA properties computation}
		    \State {Cumulative proximity:} $q_{N}(\alpha_{k}) \gets \sum_{j=1}^{N}\delta^{2}_{k,j}; $
		    \State {Discrete local density:} $D_{N}(\alpha_{k}) \gets \frac{\sum_{j=1}^{N}q_{N}(\alpha_{j})}{2Nq_{N}(\alpha_{k})}$
		    \State {Discrete typicality:} $\tau_{k}(\alpha_{k}) \gets \frac{D_{N}(\alpha_{k})}{\sum_{j=1}^{N}D_{N}(\alpha_{j})}$
		    \State {Global typicality:} $\mathbf{\tau}_{N}^{D*} \gets max(\mathbf{\tau}_{i}^{D}) \hspace{0.5cm} i = 1,\dots,N$
	    \State \Return $\mathbf{\tau}_{N}^{*}$ $\rightarrow$ $\mathbf{\alpha}(\mathbf{\tau}_{N}^{*})$
		\end{algorithmic}
	\label{alg:TDA}
	\vspace{-0.10cm}
\end{algorithm}
The \textit{typicality} property is exclusively computed by using data and has the following common properties in commonality with pdf:  \textit{i)} $0 \leq \tau_{N}(\alpha(k)) < 1$; \textit{ii)} $\sum_{j=1}^{N}\tau_{N}(\alpha(k)) = 1$. Since it is constructed from the data unlike traditional pdf, then it will not generate values of $\tau_{N}$ for infeasible virtual data points (like  over-frequency norms for a generation trip disturbance). The distribution of typicalities will be exact, unlike traditional pdf. As pdf, the higher the value of $\tau_{N}(\alpha(k))$, this is analogous to the probability of the realization in pdf, the closer it is to the mean of the distribution. In our case, this means that the bus with the highest $\tau_{N}(\alpha(k))$ is termed $\tau^{*}(k=pb)$ and is the closest to the COI, that is, its frequency $f_{k}(t)$ is the closest to $f_{coi}(t)$, and thus, is the pilot-bus of the Region.

\subsection{Regional Inertia Estimation}\label{ssec:FundRegionalH}

The inertial frequency response for a single synchronous machine is characterized by the classical swing equation, as follows~\cite{chow16}:

\begin{equation}
    \frac{d\omega(t)}{dt}=\frac{1}{2H}(P_{m}(t)-P_{e}(t) - D\Delta f(t))
    \label{eq:swingomega}
\end{equation}

\noindent where $H$ is the inertia constant of the generator, which represents the machines rotor kinetic energy in seconds at the machines rated power, $P_{m}(t)$ is the machine mechanical power provided by the primary energy source, $P_{e}(t)$ is the machine electrical power output injected at the grid, $D$ is the load damping coefficient and $\frac{d\omega(t)}{dt}$ is the generator rotor speed derivative after disturbance. Assuming that the electrical frequency $f(t)$ at the machine point of connection is approximately equal to the rotor speed and there is no reasonable time for the machine speed governor to take action during the period of the inertial response, \eqref{eq:swingomega} can be rewritten as:

\begin{equation}
    \frac{df}{dt} = \frac{-(\Delta P_{e}(t) + D\Delta f(t))}{2H}
    \label{eq:swingPe}
\end{equation}

\noindent where $\Delta P_{e}(t)$ is the amount of electrical power deviation caused by a given disturbance at the generator. If the Region is strongly connected, this representation can be extended to a whole Region. Then, an equivalent frequency of the Region ($f_{coi}(t)$) can be determined, being the COI frequency. In turn, it is an weighted average of the frequency of generators, synchronous condensers and motors by their respective inertias:

\begin{equation}
    f_{coi}(t) = \frac{\sum_{n=1}^{N_g}f_{n}(t)H_{n} + \sum_{m=1}^{N_m}f_{m}(t)H_{m}}{\sum_{n=1}^{N_g}H_{n} + \sum_{m=1}^{N_m}H_{m}}
    \label{eq:fcoi}
\end{equation}


Then, for any given disturbance in the system, like generation trip or transmission line disconnection, the equivalent inertial response of the Region can be given by the COI frequency response as follows

\begin{equation}
    \frac{df_{coi}(t)}{dt} = \frac{-(\sum_{n=1}^{N_g}\Delta P_{e}^{n}(t) + \sum_{m=1}^{N_m}\Delta P_{e}^{m}(t) + D\Delta f_{coi}(t))}{2H_{Req}}
    \label{eq:regionalInertiaResp}
\end{equation}

Now, $\Delta P_e^{n}(t)$ is the electrical power deviation of each generator and synchronous condenser connected in the Region $R$ and $\Delta P_e^{m}(t)$ is the electrical power deviation at the transmission bus connected to a relevant portion of motors.

The definition of the Region $R$, where the inertia estimation is carried out, depends on several factors such as location of the disturbance, size of the disturbance, topology of the system and exchanges at tie-lines~\cite{chow13Coherency}. It is usually performed using coherency analysis of electromechanical modes as in~\cite{Lugnani2021TDA}. 
However, due to a power system is usually very well connected within itself with weaker links to other systems, and the inertial response excites slower electromechanical loads, it is reasonable to assume that this system is a coherent Region for inertial response purposes.

In the next section, we present the methodology to estimate the regional inertial response using only synchrophasor signals from the pilot-bus. 
To this end, a parametric approach is adopted to identify an equivalent machine that encompasses the dynamics of the Region.

\section{Methodology: ARMAX-based regional inertia estimation}
\label{sec:method}

\begin{figure*}[!t]
\vspace{-1cm}
    \centering
    \includegraphics[width=\textwidth]{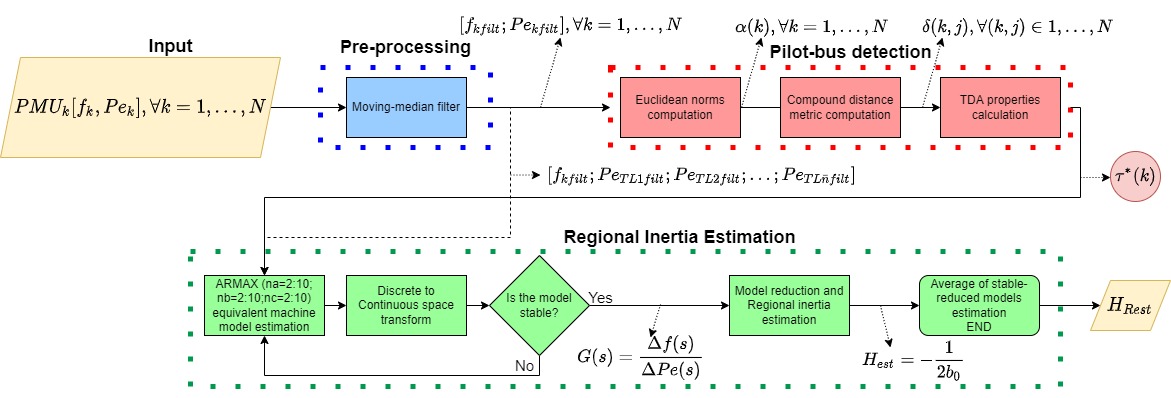}
    \caption{Pilot-bus detection and regional inertia estimation.}
    \label{fig:methodology}
\end{figure*}

The purpose of defining a pilot-bus for a Region of interest is estimating the inertia of that Region with minimal data, particularly without any model parametric information. In this section, we present the steps in the pre-processing stage for signal filtering of the data, the pilot-bus detection stage and the steps for estimation using Auto-Regressive Moving Average eXogenous input (ARMAX) model identification technique~\cite{lugnani2020armax,BrunoARMAX}. It is important to indicate that the filtered signals $f_{filt}(t)$ and $Pe_{filt}(t)$ are first used for pilot-bus detection, and then the filtered frequency signals of the pilot-bus $f_{filtPB}$ and the active power of the Region interconnections signal $Pe_{filtTL}$ are used by ARMAX for identifying the equivalent machine model for regional inertia estimation. Figure~\ref{fig:methodology} presents the overall pathway of the proposed methodology for regional inertia estimation.

\subsection{Signals pre-processing}


These are acquired by PMUs and must be filtered due to presence of non-electromechanical phenomena in the voltage phase angles that are used for frequency estimation. These phenomena come from the voltage regulation action that has no relationship to disturbances associated with frequency stability. To deal with noisy signals (and any other high-frequency noises contained in the signals), a low-pass frequency \textit{moving median} filter is applied using Matlab function \textit{movmedian} with a 5-sample window.

\subsection{Pilot-bus detection}\label{ssec:pb}

Once the signals are filtered, the TDA strategy is applied according to Algorithm~\ref{alg:TDA} from~\ref{ssec:FundPilotBus}. The filtered signals of active power and frequency and output the detected pilot-bus are the TDA inputs. It is important to emphasize some points regarding the application of TDA for pilot-bus detection: (\textit{i}) each bus $k$ will represent a point $\alpha(k)$ in the data distribution with an active power ($Pe_{k}(t)$) and a frequency ($f_{k}(t)$) component; (\textit{ii}) both power and frequency signals will be represented by the Euclidean norm respective to every other bus $j$ (as in ~\eqref{eq:EuclideanNormf} and~\eqref{eq:EuclideanNormP}), which is calculated for a typical inertial response interval of two seconds; and (\textit{iii}) the compound distance metric $\delta$ calculated between every two points has equal weight for both metrics (Line 6 of Algorithm~\ref{alg:TDA}).

With the distances, the TDA properties are calculated according to Algorithm~\ref{alg:TDA}. Where the final property, typicality $\tau_{N}(\alpha(k))$ (Line 12 of Algorithm~\ref{alg:TDA}), of each data point is calculated representing a data-driven pdf of the inertial response. The distribution of inertial responses has as mean $f_{coi}(t)$ in most cases virtual. 
In our case, the TDA renders a vector of $\tau_{N}(\alpha(k))$ values of equal dimension as the number of PMUs, where each value of $\tau_{N}$ represents the probability of a realization assuming that particular value of $\alpha(k)$. Hence, the highest value of $\tau_{N}$, that is $\tau_{N}^{*}$, is the most probable realization, which is the closest to the mean of distribution. Thereby, once all typicalities are calculated, we can detect the pilot-bus ($Pb$) as the bus corresponding to $\tau_{N}^{*}$ (Line 13 of Algorithm~\ref{alg:TDA}).

It is also important to reiterate that data-driven methods are usually event specific, so the detected pilot-bus will be valid for the particular event. However, as availability of data from WAMS is abundant, the method can be readily applied to every new event and a statistics analysis can be performed on the behavior of the pilot-bus movement according to disturbance location, size, operation point of the system and season effects on renewable generation. With the detected pilot-bus and the measurements of active power of the $\bar{n}$ interconnections of the Region ($Pe_{TL\bar{n}-filt}$), the regional equivalent inertia can be estimated using ARMAX model identification method presented in the following.

\subsection{Regional inertia estimation}\label{ssec:armax}

Once the pilot-bus is detected, the COI is also identified. Then, the regional inertial response in \eqref{eq:regionalInertiaResp} can be represented in p.u., since the active power deviation in generators and motors is approximately proportional to the active power deviation at interconnection buses among regions~\cite{Wilson18}. Additionally, since the frequency response of the pilot-bus ($f_{pb}$) is approximately the frequency response of the COI ($f_{pb} \approx f_{coi}$), thus ~\eqref{eq:regionalInertiaResp} can be rewritten as:


\begin{equation}
    \frac{df_{pb}(t)}{dt} = \frac{-(\sum_{k=1}^{\bar{n}}\Delta Pe_{TLk-filt}(t) +D\Delta f_{pb}(t))}{2H_{Req}}
    \label{eq:inertiaRespTL}
\end{equation}

\noindent where the boundary deviation is defined as $\Delta Pe_{B}(t)=\sum_{k=1}^{\bar{n}}\Delta Pe_{TLk-filt}(t)$. By taking the Laplace transform of~\eqref{eq:inertiaRespTL}, the frequency-domain inertial response can be defined by a first order transfer function, where active power deviation is an input and frequency deviation is an output, such that:

\begin{equation}
    G(s)=\frac{\Delta f_{pb}(s)}{\Delta Pe_{B}(s)} = \frac{-1/2H_{Req}}{s + D/2H_{Req}}
    \label{eq:inertiaTF}
\end{equation}

Thus, the inertial response of the selected Region is represented by the transfer function in ~\eqref{eq:inertiaTF}, but only using pilot-bus frequency measurements and interconnection buses active power deviations. To perform such assessment, an ARMAX approach is advocated~\cite{lugnani2020armax}.

To prevent outliers and gross errors, the ARMAX model estimation is performed for different orders of  polynomials such as: $A, n_{a}=[2,\dots,10]$, $B, n_{b}=[2,\dots,10]$; and $C, n_{b}=[2,\dots,10]$. 
This is carried out in a two-step manner: (\textit{i}) stability of $G_{e}(s)$, where all of transfer function  poles are analyzed; and (\textit{ii}) quality of prediction, where the normalized root squared error (NRSE) given by \eqref{eq:nrse} is determined.

\begin{equation}
    NRSE = \Bigg( 1-\frac{||f_{pb}(t)-f_{est}(t)||}{||f_{pb}(t)-mean(f_{pb}(t))||} \Bigg) \times 100 [\%]
    \label{eq:nrse}
\end{equation}

Stable models are reduced to first-order transfer functions using MATLAB function \textit{balred}. For the assessment of the inertial response, the estimated transfer function $G_{e}(s)$ in \eqref{eq:armaxTF} is inspected.

\begin{equation}
    G_{e}(s) = \frac{b_{0}}{s + a_{0}}
    \label{eq:armaxTF}
\end{equation}

Then, ~\eqref{eq:inertiaTF} is compared to infer the regional inertia as:

\begin{equation}
    H_{Req}=\frac{-1}{2b_{0}}
\end{equation}

Finally, the last step is the average of the adequate estimates of $H_{Req}$, i.e. the estimations whose $G_{e}(s)$ transfer functions are stable and whose NRSE prediction error is under 5$\%$, rendering the final estimation of the regional inertia by this exclusively data-driven method.

\section{IEEE Test Benchmark System}\label{sec:system}

The well-known IEEE 68-bus system \cite{pal2006robust} is a reduced order equivalent of the inter-connected New England test system (NETS) and New York power system (NYPS). It is composed of 16-machines represented by a sixth order model equipped with AVRs,  PSSs (PSS1a simplified with three lead-lag steps), and a generic model of governor with one operating mode representing steam turbine generator~\cite{SpeedRegIEEE2013}. The load model is represented by constant impedance. 




The contribution of induction motors to power system inertial response is considered in this work. To this end, a 10 \% of the load at each bus is represented by a dynamic load corresponding to a set of aggregated motors with an equivalent inertia of $H_{m}=5s$ (100 MVA) \cite{rahim1987aggregation,dattaray2017impact}.
Table ~\ref{tab:motors} summarizes this modification for each Region, showing the number of load buses,  the total contribution of aggregated inertia, and the ratio $\rho$ of $H_{m}/H_{g}$.


\begin{table}[!t]
    \centering
    \caption{Dynamic load for of NETS/NYPS system.}
    \begin{tabular}{cccc}
    \hline
    \hline
        Region & $\#$ buses & Region Load inertia [s] & $\rho$ \\
    \hline
    \hline
        NETS & 17 & 85 & 0.3014\\
    \hline
        NYPS & 15 & 75 & 0.1157\\
    \hline
    \hline
    \end{tabular}
    \label{tab:motors}
\end{table}

\section{Performance of the COI and Regional Inertia Estimation with Load Contribution}
\label{sec:results}

To assess the performance of the fully data-driven methodology in finding the COI and estimating the regional inertia, nonlinear time-domain simulations are performed using the ANATEM simulation software~\cite{ANAREDE17} with a total time of 20 s. Since our methodology adopts the disturbance approach according to \cite{Ashton15,wilson2019measuring,Zografos17,alshahrestani2018wams,schiffer2019online,phurailatpam2019measurement,yang2020data,makolo2021online,kerdphol2021determining}, all simulations include disturbances provoked by load steps occurring at 0.6s at the higher load buses. Active power and frequency measurements are collected using 60 phasors per second in fulfillment with the synchrophasor standard ~\cite{IEEE2011PMUstandard}.



\subsection{Application of the TDA method for pilot-bus detection}

\subsubsection{S1.w - Detection of pilot-bus per Region without motor contributions}\label{sssec:s1w}

\begin{figure}[!h]
    \centering
    \includegraphics[width=\columnwidth]{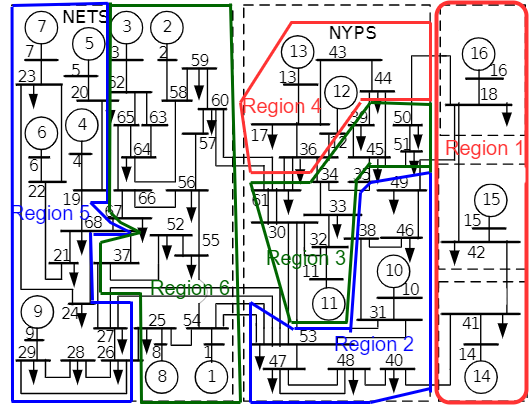}
    \caption{IEEE 68-Bus with 6 areas determined by the TDA method.}
    \label{fig:68busTDA}
\end{figure}

The application of the TDA method, described in \cite{chow13Coherency}, for the identification of coherent regions for the selected disturbance is illustrated in  Figure~\ref{fig:68busTDA}. Table~\ref{tab:s1Regions.w} shows in detail the limits of each Region, given the disturbance at bus 17.

\begin{table}[!h]
    \centering
    \caption{Regions and their respective buses for Figure~\ref{fig:68busTDA}. }
    \scalebox{0.9}{
    \begin{tabular}{cc}
    \hline
    \hline
        Region & Buses   \\
    \hline
    \hline
        1 & 14-16;18;41-42 \\
    \hline
        2 & 10;31;38;40;46-49;53  \\
    \hline
        3 & 11;30;32-35;45;50-51;61  \\
    \hline
        4 & 12-13;17;36;39;43-44  \\
    \hline
        5 & 4-7;9;19-24;26-29;68  \\
    \hline
        6 & 1-3;8;25;37;52;54-60;62-67 \\
    \hline
    \hline
    \end{tabular}
    }
    \label{tab:s1Regions.w}
\end{table}

For each Region a  pilot-bus ($Pb$) is  detected and the results are summarized in Table~\ref{tab:s1.w} along with the normalized root mean square error (NRMSE) (\%) in reference to the true calculated COI frequency response, given by:

\begin{equation}
    NRMSE = \frac{\sqrt{\frac{\sum_{t=t_{0}}^{t_{f}}(f_{pb}(t)-f_{coi}(t))^{2}}{T}}}{\bar{f}_{coi}}
\end{equation}

\noindent where $t_{0}$ is the moment of disturbance, $t_{f}$ is the two seconds assumed for the inertial response, $f_{pb}$ and $f_{coi}$ are the detected pilot-bus frequency response and the COI calculated frequency response using knowledge of the model, respectively, $T$ is the number of samples of the window, and $\bar{f}_{coi}$ is the mean of the frequency response of the COI. Besides the NRMSE calulated to the pilot-bus frequency response ($f_{pb}$), Table~\ref{tab:s1.w} also shows the NRMSE  calculated for the average of the generators frequency response ($f_{g}$), and the average frequency response for all buses ($f_{b}$), all in reference to $f_{coi}$. Table~\ref{tab:s1.w} also provides the NRMSE threshold of first quartile ($1^{st}-q$) of the frequency response of all buses in the Region.

\begin{table}[!h]
    \centering
    \caption{Regions, detected pilot-buses and NRMSE of frequency responses, with respect to $f_{coi}$ for S1.w.}
    \scalebox{0.9}{
    \begin{tabular}{cccccc}
    \hline
    \hline
        Region & $Pb$ & $f_{pb}$ & $f_{g}$ & $f_{b}$ &  $1^{st}-q$  \\
    \hline
    \hline
        1 & 41 & 2.12 & 0.62 & 0.61 & 2.11 \\
    \hline
        2 & 48 & 0 & 0 & 2.13 & 0.88 \\
    \hline
        3 & 35 & 2.62 & 0 & 1.96 & 1.18 \\
    \hline
        4 & 13 & 6.95 & 2.06 & 3.97 & 2.48 \\
    \hline
        5 & 22 & 0.50 & 0.05 & 0.69 & 0.71 \\
    \hline
        6 & 37 & 0.70 & 0.14 & 0.97 & 1.08 \\
    \hline
    \hline
    \end{tabular}
    }
    \label{tab:s1.w}
\end{table}



For regions 2 and 3 (NYPS, except Region of disturbance), the attained results are within the $1^{st}-q$ of the distribution, without surpassing the average response of the generators frequency response $f_{g}$ and the average frequency response of all buses in the Region $f_{b}$. This is because regions 2 and 3 are each composed of one generator, thereby the frequency response of the generator is equivalent to the COI frequency response ($f_{g}=f_{coi}$). Also, these regions are closer to the disturbance and with a smaller sample of buses, impoverishing the detection. 
For Region 4, the results were impaired due to Region having only two generators with high inertias each, that is, the inertia of generator 12 $H_{g12}=92.3 s$ and generator 13 $H_{g13}=496 s$. The TDA method detected that generator 13 had a higher influence in the Region COI frequency response by pointing its connection bus as the pilot-bus. However, since generator 12 corresponds to 15.7$\%$ of the inertia of the Region, the error is increased. We can note that for regions with better evenly distributed inertia, the results of the pilot-bus become more precise. For regions 5 and 6 (NETS), the detected pilot-buses also present results within the the first quartile $1^{st}-q$, but this time, the TDA pilot-bus frequency response ($f_{pb}$) is better than the average of the inertial response of all buses in the Region. Nevertheless, the inertia values of generators in the NETS vary only slightly, thus the weights of their inertias are similar to every generator, hence the generators frequency response ($f_{g}$) is closer to the COI frequency response ($f_{coi}$) than ($f_{pb}$). 


Additional tests were carried out introducing step changes of 10\% in the 5 largest loads of both NYPS and NETS systems, showing similar results, confirming the validity of the TDA methodology. However, a more realistic scenario includes the contribution of the load inertial response, which is particularly relevant in today's power systems with high penetration of CCG and lowering generator inertia contribution. With this more realistic scenario we show that considering only the mean of the inertial responses of synchronous generators may be a poorer choice of representation of the $f_{coi}$.

\subsubsection{S1.m - Detection of pilot-bus per Region including motors}\label{sssec:s1m}

\begin{figure}[!b]
    \centering
    \includegraphics[width=\columnwidth]{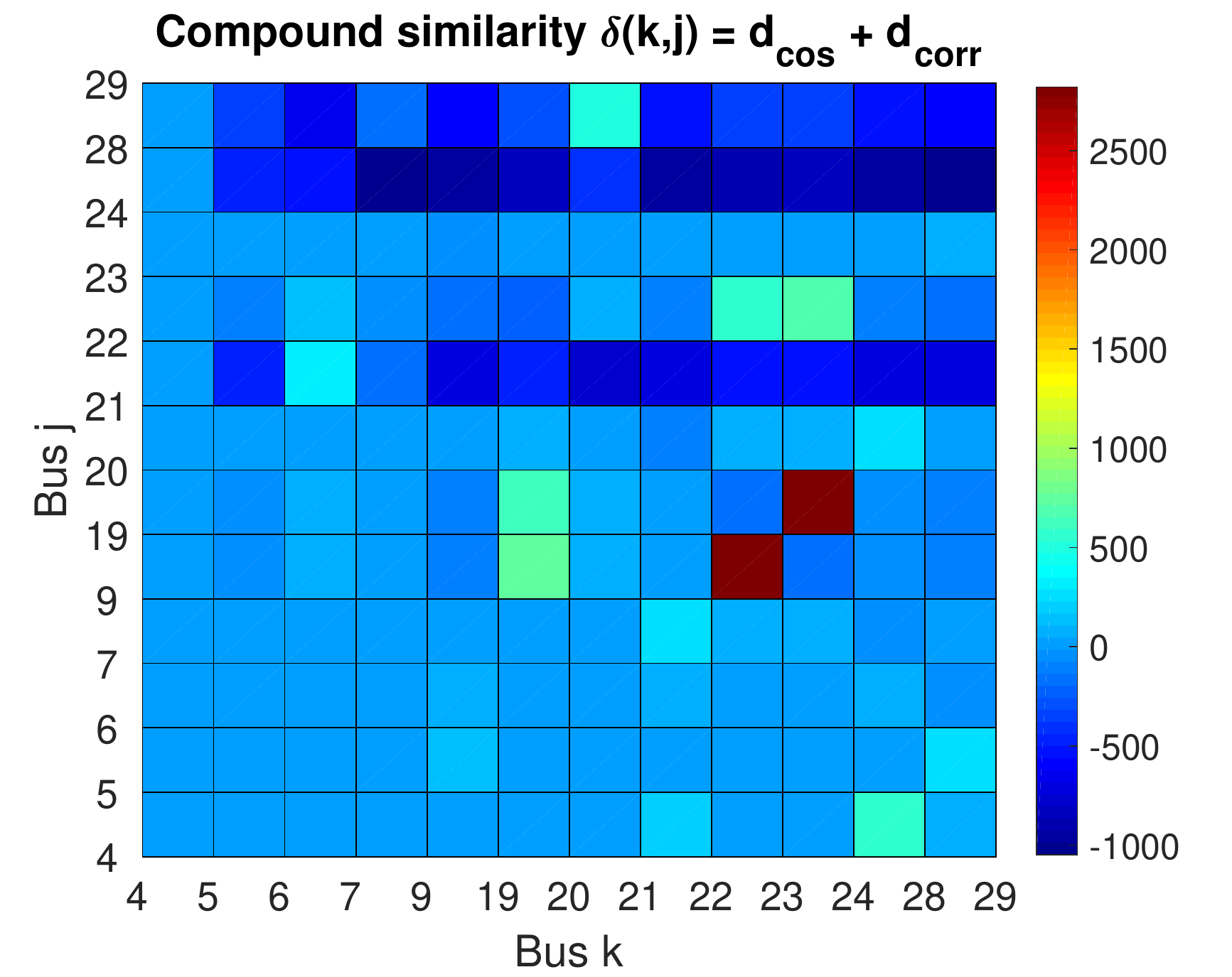}
    \caption{Metrics Region 5.}
    \label{fig:DistGrp5_S1m}
\end{figure}

The load's inertia is relevant and must not be ignored for pilot-bus detection; however, this configuration essentially displaces the COI's position. Thus, the use of generators mean frequency response $f_{g}$ as pilot-bus becomes inaccurate. Therefor, this scenario includes the same load step (10\%) over the dynamical loads added to the system. The TDA method is applied to the previously identified regions for the detection of the pilot-bus. As an example, Fig.~\ref{fig:DistGrp5_S1m} shows the calculated similarities for Region 5.

\begin{figure}[!t]
\vspace{-1cm}
    \centering
    \includegraphics[width=\columnwidth]{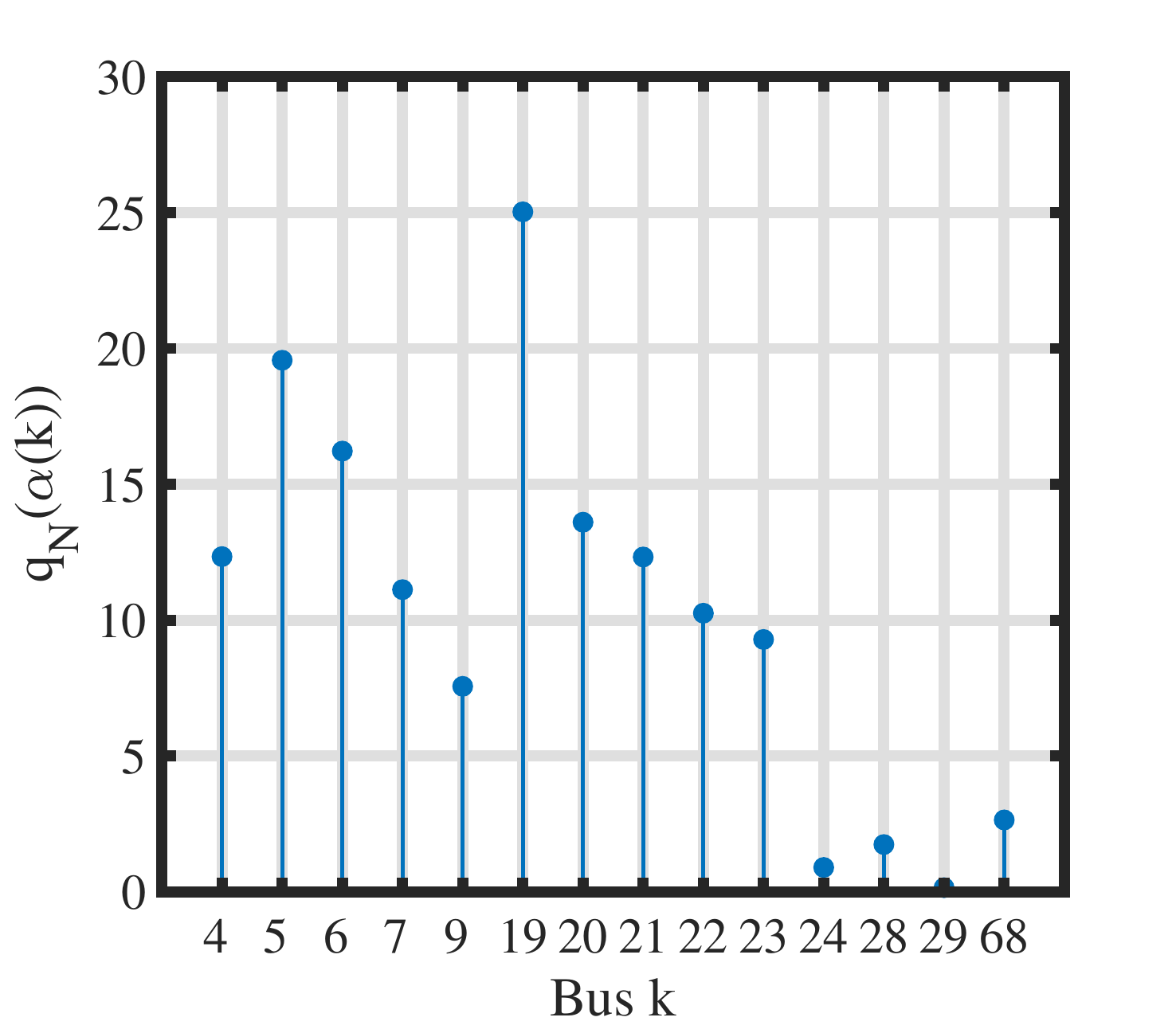}
    \caption{Cumulative proximities Region 5.}
    \label{fig:QnGrp5_S1m}
\end{figure}

Notice that the values in Fig.~\ref{fig:DistGrp5_S1m} range in an arbitrary interval, containing positive and negative values. The first steps in the TDA method, i.e. the computation of the cumulative proximity $q_{N}(\alpha(k))$ and the standardized eccentricity $\epsilon_{N}(\alpha(k))$, deal with the normalization of the data, like other methods. But in our proposal, the normalization process does not assume any model of distribution for the collected data, but rather uses the data exclusively. This generates only feasible values in the normalized range, i.e. no compound distance produced by unrealistic frequency deviation or active power deviation values would be part of the range. Fig.~\ref{fig:QnGrp5_S1m} displays the calculated proximities for the same Region.

\begin{figure*}[!h]
    \centering
    \includegraphics[trim={2cm 0.1cm 4.5cm 0.1cm},clip,width=\textwidth]{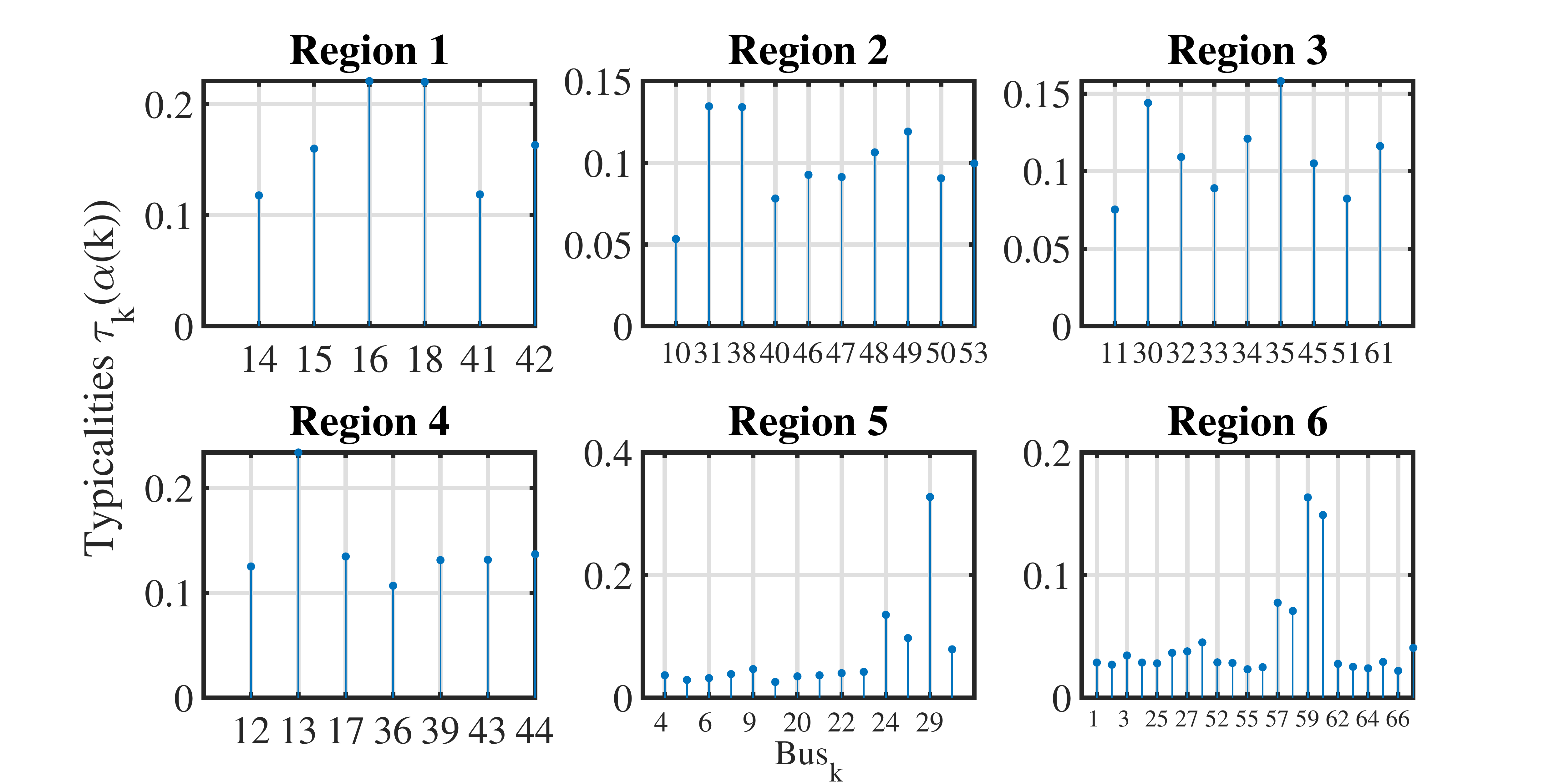}
    \caption{Typicalities for S1.m.}
    \label{fig:S1m}
\end{figure*}

The detected pilot-bus with the corresponding NRMSE is presented in Tab.~\ref{tab:s1.m}, showing the displacement of the COI when load inertial response is added to the system, as the results from the TDA pilot-bus frequency response $f_{pb}$ become equivalent or more precise, i.e. with a lower RMSE, than the generators mean frequency response $f_{g}$. The typicality distribution of each Region is presented in Fig.~\ref{fig:S1m}, showing that even though the representation of generators is significant, the load buses dislocate the mean of the distribution, having themselves more participation in the inertial frequency response. 

\begin{table}[!h]
    \centering
    \caption{Regions, detected pilot-buses and NRMSE of frequency responses, with respect to $f_{coi}$ for S1.m.}
    \scalebox{0.9}{
    \begin{tabular}{cccccc}
    \hline
    \hline
        Region & $Pb$ & $f_{pb}$ & $f_{g}$ & $f_{b}$ & $1^{st}-q$  \\
    \hline
    \hline
        1 & 41 & 1.27 & 0.21 & 0.22 & 1.22 \\
    \hline
        2 & 31 & 0 & 0.17 & 1.64 & 1.79 \\
    \hline
        3 & 35 & 1.44 & 1.44 & 3.29 & 1.96 \\
    \hline
        4 & 13 & 3.42 & 3.14 & 4.57 & 3.58 \\
    \hline
        5 & 29 & 0.22 & 0.31 & 0.49 & 0.49 \\
    \hline
        6 & 59 & 0.19 & 0.25 & 0.19 & 0.28 \\
    \hline
    \hline
    \end{tabular}
    }
    \label{tab:s1.m}
\end{table}

As expected from Tab.~\ref{tab:s1.m}, it is noteworthy to validate that: the presumption that $f_{g}$ is the best approximation of the COI is erroneous, since the load is not represented by constant impedances. Then, the TDA matches the result from $f_{g}$ for regions 2 and 3 due to the $f_{pb}$ frequency response surpassing the inertial response of the single generation in Region 2, as the TDA pilot-bus detection takes into account load contribution. For Region 3, the RMSE of the bus detected by the TDA method coincides with the error of the Region $f_{g}$.

For Region 4, the result of the pilot-bus detected by the TDA method approximates $f_{g}$ as the contribution of load inertia is small in this Region, compared with the generator's inertia.  
For regions 5 and 6, the RMSE of $f_{pb}$ is smaller than $f_{g}$, as these regions have a high number of load buses and as we assume equal distribution of dynamical loads among load buses, the value of $f_{coi}$ displaces more from the weighted mean of the generators.

\begin{figure}[!h]
    \centering
    \includegraphics[trim={2.5cm 0.2cm 3.5cm 0.3cm},clip,width=\columnwidth]{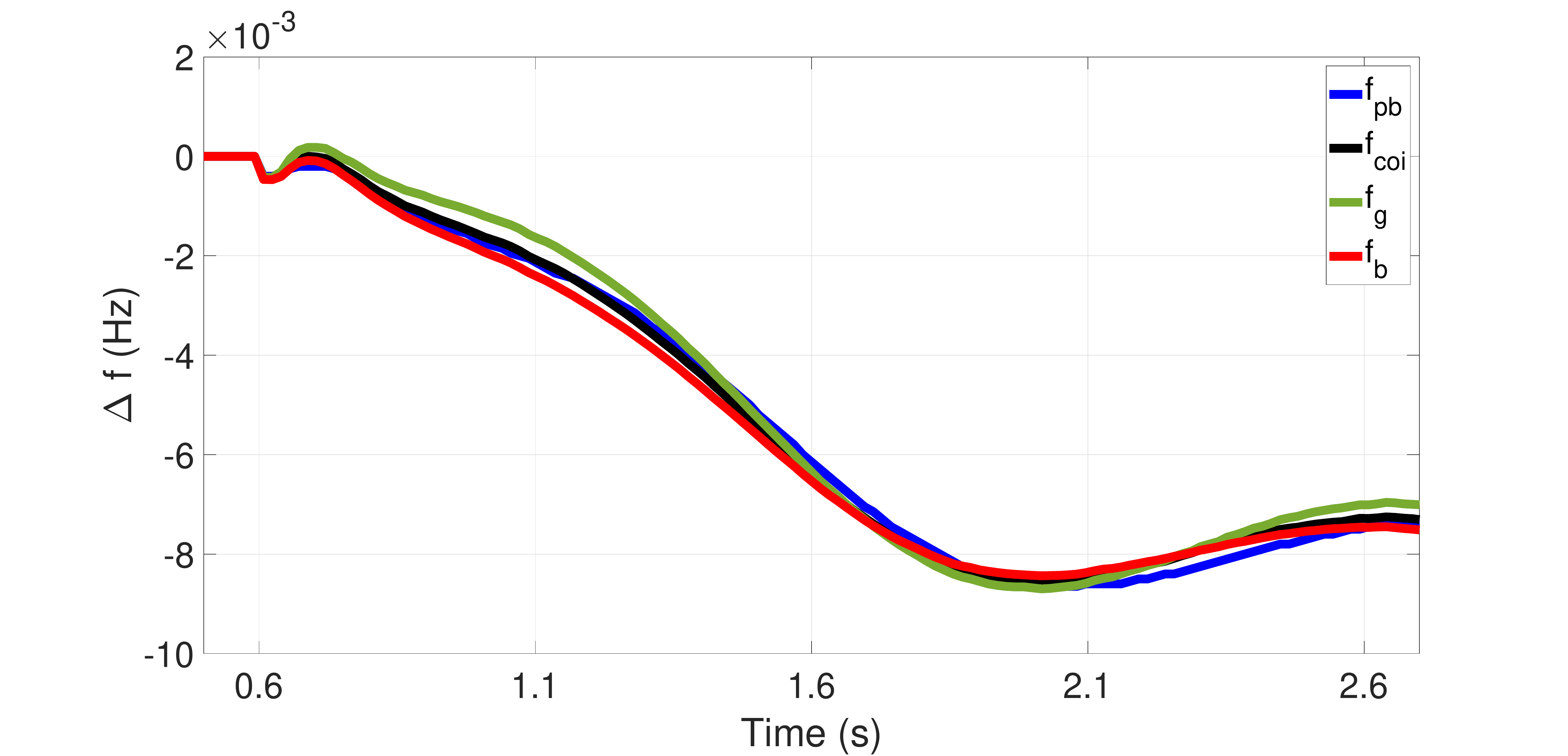}
    \caption{Frequency response comparison of $f_{coi}$, $f_{pb}$, $f_{g}$ and $f_{b}$ for Region 5.}
    \label{fig:ComparisonRegion5m}
\end{figure}

It is also noteworthy to remark that the pilot-bus detected by the TDA method in all five regions is within the first quartile, indicating a consistency in approximating the distribution of data with the proposed method. For instance, selecting the pilot-bus and the mean of generators and every bus for Region 5, Fig.~\ref{fig:ComparisonRegion5m} illustrates the comparison of the frequency responses of the COI. Where it is noticed that the pilot-bus frequency response $f_{pb}$ properly tracks the trajectory of the COI frequency response $f_{coi}$, even though no model information is provided for the TDA method. Additionally, the displacement  of the COI is evident, as the frequency response of the generators $f_{g}$ becomes more distant from the COI.

Next, we will apply the equivalent inertia's estimation method per Region for the above cases.

\subsection{Regional inertia estimation using the detected pilot-bus}

In this section, the methodology schematized in Fig. \ref{fig:methodology} is followed taking advantage of the right detection of the pilot-bus provided by the TDA features, according to Section \ref{ssec:pb}, to estimate the regional inertia seen from the COI or pilot-bus. This assessment is accomplished thanks to an ARMAX-based identification approach that seeks representing the total inertia per area as the inertia of an equivalent machine, such that this machine encompasses the regional dynamic. Finally, the application is carried out employing the cases described above, i.e. without and with motor inertial contributions. 


\subsubsection{ARMAX Regional inertia $H_{est}$ estimation for case~\ref{sssec:s1w}}

Once, the ARMAX-based methodology is applied according to Section \ref{ssec:armax}, the assessment of the regional inertia is achieved and compared with the reference values used for simulations. Where $H_{ref}$ in Table ~\ref{tab:s1.wHreg} represents the sum of the total inertia per Region, $H_est$ denotes the estimate by our proposal, and $RE$ is the relative error in percentage. To produce the regional inertia estimation, the interconnections' active power deviation of each Region are used as input signals, that is $u(t)= \Delta Pe_{B}(t) = \sum \Delta Pe_{TL(t)}$; meanwhile the pilot-bus frequency signal deviation  is employed as output signal $y(t)$. Then, all steps contained in the green dotted box in Fig.~\ref{fig:methodology} are applied to these signals. The ARMAX model is estimated for equal orders of $[n_a,n_{b},n_{c}]=[2,\dots,10]$, and the final inertia estimation is considered as the average of all accepted estimates. 


From Table~\ref{tab:s1.wHreg}, it is noteworthy to remark that the assessment of the COI together with the regional inertia results in errors in line with those found in the literature for regional estimation, even for pilot-buses with greater error to the true COI frequency response than the average of generators frequencies. 


\begin{table}[!h]
    \centering
    \caption{Regional Inertia Estimates for Case ~\ref{sssec:s1w}}
    \scalebox{0.9}{
    \begin{tabular}{cccc}
    \hline
    \hline
        Region & $H_{ref}$ & $H_{est}$ & RE [$\%$] \\
    \hline
    \hline
        1 & 1050 & 1777.4 & 2.61 \\
    \hline
        2 & 31 & 30.4 & 1.94 \\
    \hline
        3 & 28.2 & 28.9 & 2.48  \\
    \hline
        4 & 588.3 & 568.9 & 3.29  \\
    \hline
        5 & 115.9 & 119.4 & 3.02  \\
    \hline
        6 & 106.7 & 110.3 & 3.37  \\
    \hline
    \hline
    \end{tabular}
    }
    \label{tab:s1.wHreg}
\end{table}

\subsubsection{ARMAX Regional inertia $H_{est}$ estimation for case~\ref{sssec:s1m}}

\begin{figure}[!h]
    \centering
    \includegraphics[trim={2.5cm 0.2cm 3.5cm 0.3cm},clip,width=\columnwidth]{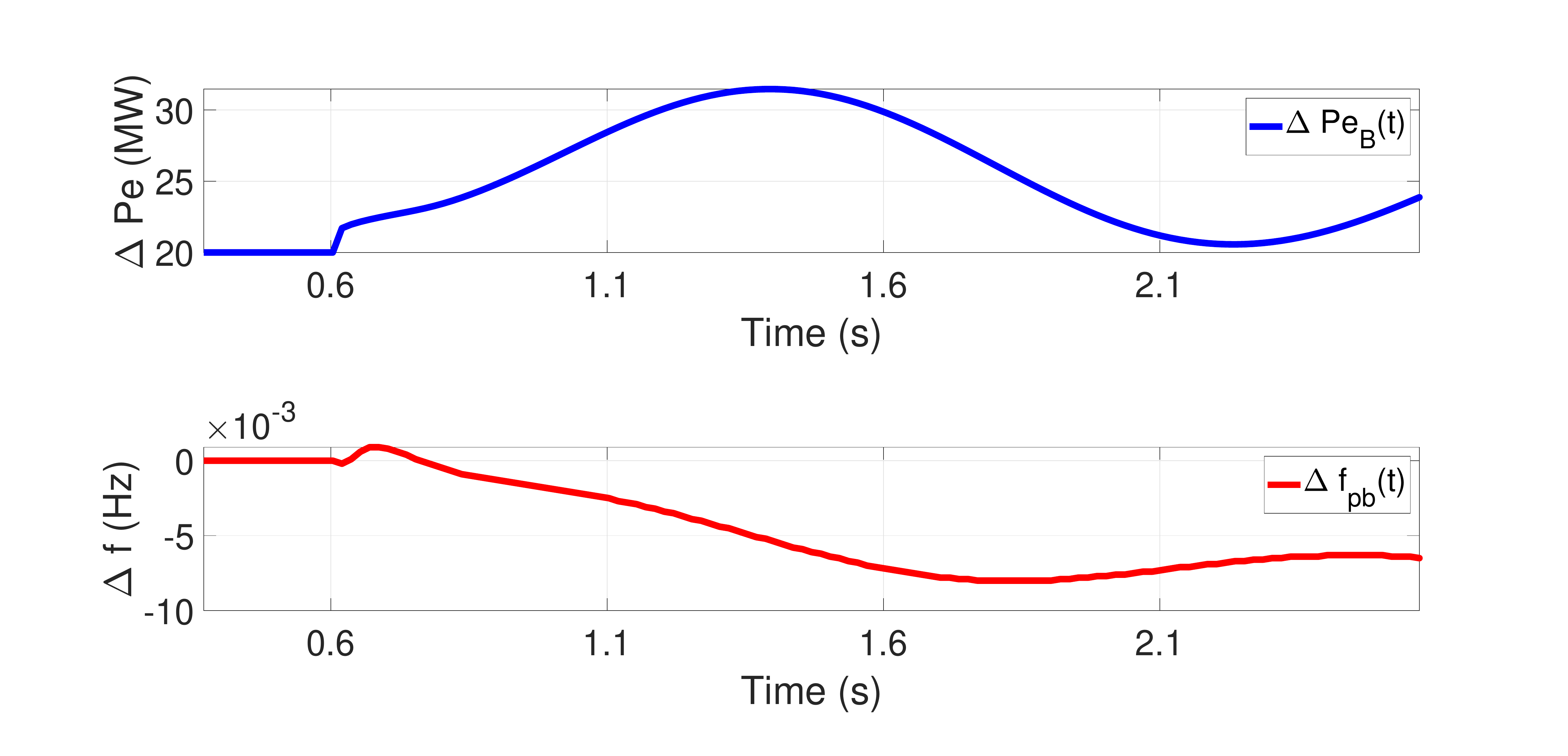}
    \caption{Input signal - active power of Region 5 tie-lines. Output signal - pilot-bus frequency deviation of Region 5}
    \label{fig:InputOutputRegion5}
\end{figure}

Here, all areas (except Area 1) have been added with the inertia provided by motors, resulting in the estimated regional inertias summarized in Table~\ref{tab:s1.mHreg}. Note that the additional inertia provided by the motors is also considered as reference in $H_{ref}$. This is an important aspect of regional inertia that is not usually considered in most methods found in literature~\cite{Ashton15,alshahrestani2018wams} and applied to real systems, i.e. the ability of estimation methods in capturing load inertial contribution. This proposal provides a reference estimation of regional inertia with load contribution ($H_{est}$) with reasonable errors ($RE$), so applications to real systems can quantify the contributions of their respective loads. Our proposed approach does not require any model information, neither simplifies the COI inertial response $f_{coi}$ by the average of generators, disregarding other sources of inertial response. Figure~\ref{fig:InputOutputRegion5} shows the input and output signals for this case, which are used by the ARMAX estimation methodology.

\begin{table}[!h]
    \centering
    \caption{Regional Inertia Estimates for Case~\ref{sssec:s1m}}
    \scalebox{0.9}{
    \begin{tabular}{cccc}
    \hline
    \hline
        Region & $H_{ref}$ & $H_{est}$ & RE [$\%$] \\
    \hline
    \hline
        1 & 1050 & 970.49 & 7.57 \\
    \hline        
        2 & 61 & 63.3 & 3.77 \\
    \hline
        3 & 58.2 & 54.3 & 6.7  \\
    \hline
        4 & 603.3 & 592.1 & 1.86  \\
    \hline
        5 & 190.3 & 183.4 & 3.63  \\
    \hline
        6 & 148 & 144.5 & 2.36  \\
    \hline
    \hline
    \end{tabular}
    }
    \label{tab:s1.mHreg}
\end{table}

From the numerical results, the ARMAX-based methodology is able to estimate the regional inertia using the pilot-bus provided by the TDA method with errors similar to those ones found in literature~\cite{wilson2019measuring,alshahrestani2018wams}, despite there are scarce resources to estimate the load contribution. 


\section{Conclusion}\label{sec:conclusion}

This work proposes an overall fully data-driven methodology for the COI and regional inertia estimation. The proposed method identifies a candidate pilot-bus, as the Center of Inertia, belonging to a Region. It estimates the inertia using only available PMU measurements such as frequency data from that bus and active power signals from interconnections of the Region after disturbances. The method is tested for the NETS/NYPS benchmark system and its modified version with dynamic load representation of aggregated induction motors to represent the inertial contribution of the load. Finally, the regional inertia is estimated providing satisfactory results.

Future works include the assessment of the COI and regional inertia using an ambient data approach. Likewise, 
the estimation of the load damping coefficient and the equivalent droop of the Region, as well as, an analysis of the system inertia distribution using the TDA method.

\bibliographystyle{IEEEtran}
\bibliography{biblio}

\begin{thebibliography}{10}
\providecommand{\url}[1]{#1}
\csname url@samestyle\endcsname
\providecommand{\newblock}{\relax}
\providecommand{\bibinfo}[2]{#2}
\providecommand{\BIBentrySTDinterwordspacing}{\spaceskip=0pt\relax}
\providecommand{\BIBentryALTinterwordstretchfactor}{4}
\providecommand{\BIBentryALTinterwordspacing}{\spaceskip=\fontdimen2\font plus
\BIBentryALTinterwordstretchfactor\fontdimen3\font minus
  \fontdimen4\font\relax}
\providecommand{\BIBforeignlanguage}[2]{{%
\expandafter\ifx\csname l@#1\endcsname\relax
\typeout{** WARNING: IEEEtran.bst: No hyphenation pattern has been}%
\typeout{** loaded for the language `#1'. Using the pattern for}%
\typeout{** the default language instead.}%
\else
\language=\csname l@#1\endcsname
\fi
#2}}
\providecommand{\BIBdecl}{\relax}
\BIBdecl

\bibitem{milano2018foundations}
F.~Milano, F.~D{\"o}rfler, G.~Hug, D.~J. Hill, and G.~Verbi{\v{c}},
  ``Foundations and challenges of low-inertia systems,'' in \emph{2018 Power
  Systems Computation Conference (PSCC)}.\hskip 1em plus 0.5em minus
  0.4em\relax IEEE, 2018, pp. 1--25.

\bibitem{miller12}
L.~{Ruttledge}, N.~W. {Miller}, J.~{O'Sullivan}, and D.~{Flynn}, ``Frequency
  response of power systems with variable speed wind turbines,'' \emph{IEEE
  Transactions on Sustainable Energy}, vol.~3, no.~4, Oct 2012.

\bibitem{arani14}
M.~F.~M. {Arani} and E.~F. {El-Saadany}, ``Implementing virtual inertia in
  {DFIG}-based wind power generation,'' \emph{IEEE Transactions on Power
  Systems}, vol.~28, no.~2, May 2013.

\bibitem{Wall14}
P.~Wall and V.~Terzija, ``Simultaneous estimation of the time of disturbance
  and inertia in power systems,'' \emph{IEEE Trans. Power Del.}, vol.~29,
  no.~4, 2014.

\bibitem{gorbunov2019estimationGen}
A.~Gorbunov, A.~Dymarsky, and J.~Bialek, ``Estimation of parameters of a
  dynamic generator model from modal pmu measurements,'' \emph{IEEE
  Transactions on Power Systems}, vol.~35, no.~1, pp. 53--62, 2019.

\bibitem{lugnani2020armax}
L.~Lugnani, D.~Dotta, C.~Lackner, and J.~Chow, ``{ARMAX}-based method for
  inertial constant estimation of generation units using synchrophasors,''
  \emph{Electric Power Systems Research}, vol. 180, p. 106097, 2020.

\bibitem{cepeda2014_realCOI1}
J.~C. Cepeda, J.~L. Rueda, D.~G. Colom{\'e}, and D.~E. Echeverr{\'\i}a,
  ``Real-time transient stability assessment based on centre-of-inertia
  estimation from phasor measurement unit records,'' \emph{IET Generation,
  Transmission \& Distribution}, vol.~8, no.~8, pp. 1363--1376, 2014.

\bibitem{milano2017rotor}
F.~Milano, ``Rotor speed-free estimation of the frequency of the center of
  inertia,'' \emph{IEEE Transactions on Power Systems}, vol.~33, no.~1, pp.
  1153--1155, 2017.

\bibitem{zhao2018robust}
J.~Zhao, Y.~Tang, and V.~Terzija, ``Robust online estimation of power system
  center of inertia frequency,'' \emph{IEEE Transactions on Power Systems},
  vol.~34, no.~1, pp. 821--825, 2018.

\bibitem{you2020calculate}
S.~You, H.~Li, S.~Liu, K.~Sun, W.~Wang, W.~Qiu, and Y.~Liu, ``Calculate
  center-of-inertia frequency and system rocof using pmu data,'' \emph{arXiv
  preprint arXiv:2010.12381}, 2020.

\bibitem{azizi2020local}
S.~Azizi, M.~Sun, G.~Liu, and V.~Terzija, ``Local frequency-based estimation of
  the rate of change of frequency of the center of inertia,'' \emph{IEEE
  Transactions on Power Systems}, vol.~35, no.~6, pp. 4948--4951, 2020.

\bibitem{gorbunov2022COIambient}
A.~Gorbunov, J.~C.-H. Peng, J.~W. Bialek, and P.~Vorobev, ``Can
  center-of-inertia model be identified from ambient frequency measurements,''
  \emph{IEEE Transactions on Power Systems}, 2022.

\bibitem{Ashton15}
P.~M. Ashton, C.~S. Saunders, G.~A. Taylor, A.~M. Carter, and M.~E. Bradley,
  ``Inertia estimation of the {GB} power system using synchrophasor
  measurements,'' \emph{IEEE Transactions on Power Systems}, vol.~30, no.~2,
  2015.

\bibitem{wilson2019measuring}
D.~Wilson, J.~Yu, N.~Al-Ashwal, B.~Heimisson, and V.~Terzija, ``Measuring
  effective area inertia to determine fast-acting frequency response
  requirements,'' \emph{International Journal of Electrical Power \& Energy
  Systems}, vol. 113, pp. 1--8, 2019.

\bibitem{Zografos17}
D.~Zografos, M.~Ghandhari, and K.~Paridari, ``Estimation of power system
  inertia using particle swarm optimization,'' in \emph{Intelligent System
  Application to Power Systems (ISAP), 2017 19th International Conference
  on}.\hskip 1em plus 0.5em minus 0.4em\relax IEEE, 2017.

\bibitem{alshahrestani2018wams}
A.~Alshahrestani, M.~Golshan, and H.~H. Alhelou, ``Wams based online estimation
  of total inertia constant and damping coefficient for future smart grid
  systems,'' in \emph{2018 Smart Grid Conference (SGC)}.\hskip 1em plus 0.5em
  minus 0.4em\relax IEEE, 2018, pp. 1--5.

\bibitem{schiffer2019online}
J.~Schiffer, P.~Aristidou, and R.~Ortega, ``Online estimation of power system
  inertia using dynamic regressor extension and mixing,'' \emph{IEEE
  Transactions on Power Systems}, vol.~34, no.~6, pp. 4993--5001, 2019.

\bibitem{phurailatpam2019measurement}
C.~Phurailatpam, Z.~H. Rather, B.~Bahrani, and S.~Doolla, ``Measurement-based
  estimation of inertia in ac microgrids,'' \emph{IEEE Transactions on
  Sustainable Energy}, vol.~11, no.~3, pp. 1975--1984, 2019.

\bibitem{yang2020data}
D.~Yang, B.~Wang, G.~Cai, Z.~Chen, J.~Ma, Z.~Sun, and L.~Wang, ``Data-driven
  estimation of inertia for multi-area interconnected power systems using
  dynamic mode decomposition,'' \emph{IEEE Transactions on Industrial
  Informatics}, 2020.

\bibitem{makolo2021online}
P.~Makolo, R.~Zamora, and T.-T. Lie, ``Online inertia estimation for power
  systems with high penetration of res using recursive parameters estimation,''
  \emph{IET Renewable Power Generation}, 2021.

\bibitem{kerdphol2021determining}
T.~Kerdphol, M.~Watanabe, R.~Nishikawa, T.~Tamaki, and Y.~Mitani, ``Determining
  inertia of 60 hz japan power system using pmus from power loss event,'' in
  \emph{2021 IEEE Texas Power and Energy Conference (TPEC)}.\hskip 1em plus
  0.5em minus 0.4em\relax IEEE, 2021, pp. 1--5.

\bibitem{zhang2017}
J.~Zhang and H.~Xu, ``Online identification of power system equivalent inertia
  constant,'' \emph{IEEE Transactions on Industrial Electronics}, vol.~64,
  no.~10, 2017.

\bibitem{tamrakar2020inertia}
U.~Tamrakar, N.~Guruwacharya, N.~Bhujel, F.~Wilches-Bernal, T.~M. Hansen, and
  R.~Tonkoski, ``Inertia estimation in power systems using energy storage and
  system identification techniques,'' in \emph{2020 International Symposium on
  Power Electronics, Electrical Drives, Automation and Motion (SPEEDAM)}.\hskip
  1em plus 0.5em minus 0.4em\relax IEEE, 2020, pp. 577--582.

\bibitem{Wilson18}
K.~Tuttelberg, J.~Kilter, D.~H. Wilson, and K.~Uhlen, ``Estimation of power
  system inertia from ambient wide area measurements,'' \emph{IEEE Transactions
  on Power Systems}, 2018.

\bibitem{cai2019inertia}
G.~Cai, B.~Wang, D.~Yang, Z.~Sun, and L.~Wang, ``Inertia estimation based on
  observed electromechanical oscillation response for power systems,''
  \emph{IEEE Transactions on Power Systems}, vol.~34, no.~6, pp. 4291--4299,
  2019.

\bibitem{yang2020ambient}
D.~Yang, B.~Wang, J.~Ma, Z.~Chen, G.~Cai, Z.~Sun, and L.~Wang,
  ``Ambient-data-driven modal-identification-based approach to estimate the
  inertia of an interconnected power system,'' \emph{IEEE Access}, vol.~8, pp.
  118\,799--118\,807, 2020.

\bibitem{Wang2022InertiaAmbientOscillation}
B.~Wang, D.~Yang, G.~Cai, J.~Ma, Z.~Chen, and L.~Wang, ``Online inertia
  estimation using electromechanical oscillation modal extracted from
  synchronized ambient data,'' \emph{Journal of Modern Power Systems and Clean
  Energy}, vol.~10, no.~1, pp. 241--244, 2022.

\bibitem{cui2020ambient}
Y.~Cui, S.~You, and Y.~Liu, ``Ambient synchrophasor measurement based system
  inertia estimation,'' in \emph{2020 IEEE Power \& Energy Society General
  Meeting (PESGM)}.\hskip 1em plus 0.5em minus 0.4em\relax IEEE, 2020, pp.
  1--5.

\bibitem{allella2020line}
F.~Allella, E.~Chiodo, G.~M. Giannuzzi, D.~Lauria, and F.~Mottola, ``On-line
  estimation assessment of power systems inertia with high penetration of
  renewable generation,'' \emph{IEEE Access}, vol.~8, pp. 62\,689--62\,697,
  2020.

\bibitem{zeng2020onlineCOIambientAccess}
F.~Zeng, J.~Zhang, G.~Chen, Z.~Wu, S.~Huang, and Y.~Liang, ``Online estimation
  of power system inertia constant under normal operating conditions,''
  \emph{IEEE Access}, vol.~8, pp. 101\,426--101\,436, 2020.

\bibitem{baruzzi2021analysis}
V.~Baruzzi, M.~Lodi, A.~Oliveri, and M.~Storace, ``Analysis and improvement of
  an algorithm for the online inertia estimation in power grids with res,'' in
  \emph{2021 IEEE International Symposium on Circuits and Systems
  (ISCAS)}.\hskip 1em plus 0.5em minus 0.4em\relax IEEE, 2021, pp. 1--5.

\bibitem{guo2022onlineHCOI}
J.~Guo, X.~Wang, and B.-T. Ooi, ``Online purely data-driven estimation of
  inertia and center-of-inertia frequency for power systems with vsc-interfaced
  energy sources,'' \emph{International Journal of Electrical Power \& Energy
  Systems}, vol. 137, p. 107643, 2022.

\bibitem{ash15}
M.~{Khan}, P.~M. {Ashton}, M.~{Li}, G.~A. {Taylor}, I.~{Pisica}, and J.~{Liu},
  ``Parallel detrended fluctuation analysis for fast event detection on massive
  {PMU} data,'' \emph{IEEE Transactions on Smart Grid}, vol.~6, no.~1, Jan
  2015.

\bibitem{chow13Coherency}
J.~H. Chow, \emph{Power system coherency and model reduction}.\hskip 1em plus
  0.5em minus 0.4em\relax Springer, 2013.

\bibitem{milano2016frequency}
F.~Milano and A.~Ortega, ``Frequency divider,'' \emph{IEEE Transactions on
  Power Systems}, vol.~32, no.~2, pp. 1493--1501, 2016.

\bibitem{angelov2019empirical}
P.~P. Angelov and X.~Gu, \emph{Empirical approach to machine learning}.\hskip
  1em plus 0.5em minus 0.4em\relax Springer, 2019.

\bibitem{Lugnani2021TDA}
L.~Lugnani, M.~Paternina, D.~Dotta, J.~Chow, and Y.~Liu, ``Power system
  coherency detection from wide-area measurements by typicality-based data
  analysis,'' \emph{IEEE Transactions on Power Systems}, vol.~37, no.~1, pp.
  388--401, 2022.

\bibitem{chow16}
P.~W. Sauer, M.~Pai, and J.~Chow, \emph{Power system dynamics and
  stability}.\hskip 1em plus 0.5em minus 0.4em\relax John Wiley \& Sons, 2016.

\bibitem{BrunoARMAX}
B.~Pinheiro, L.~Lugnani, and D.~Dotta, ``A procedure for the estimation of
  frequency response using a data-driven method,'' in \emph{2021 IEEE Power
  Energy Society General Meeting (PESGM)}, 2021, pp. 01--05.

\bibitem{pal2006robust}
B.~Pal and B.~Chaudhuri, \emph{Robust control in power systems}.\hskip 1em plus
  0.5em minus 0.4em\relax Springer Science \& Business Media, 2006.

\bibitem{SpeedRegIEEE2013}
R.~Boyer, L.~Hajagos, K.~Chan, L.~Hannett, G.~Chown, W.~Hofbauer, J.~Feltes,
  F.~Modau, C.~Grande-Moran, M.~Patel, L.~Gérin-Lajoie, S.~Patterson,
  F.~Langenbacher, S.~Sterpu, D.~Leonard, A.~Schneider, L.~Lima, and
  S.~Undrill, ``Dynamic models for turbine governors in power system studies,''
  IEEE Power $\&$ Energy Society, Technical Report PES-TR1, Jan. 2013.

\bibitem{rahim1987aggregation}
A.~Rahim and A.~Laldin, ``Aggregation of induction motor loads for transient
  stability studies,'' \emph{IEEE transactions on energy conversion}, no.~1,
  pp. 55--61, 1987.

\bibitem{dattaray2017impact}
P.~Dattaray, P.~Wall, V.~Terzija, P.~Mohapatra, and J.~Yu, ``Impact of location
  and composition of dynamic load on the severity of ssr in meshed power
  systems,'' in \emph{2017 IEEE Manchester PowerTech}.\hskip 1em plus 0.5em
  minus 0.4em\relax IEEE, 2017, pp. 1--6.

\bibitem{ANAREDE17}
E.~G. Department, \emph{Grid Analysis Software}, Centro de Pesquisas de Energia
  El\'{e}trica, 2017, (\textit{in Portuguese}).

\bibitem{IEEE2011PMUstandard}
``{IEEE} standard for synchrophasor data transfer for power systems,''
  \emph{{IEEE} Std C37.118.2-2011 (Revision of IEEE Std C37.118-2005)}, pp.
  1--53, 2011.

\end{thebibliography}











\newpage

 




\vfill

\end{document}